\documentclass{article}
\usepackage{amsmath,amssymb,amsfonts,graphicx,bm}
\usepackage{pgfplots}

\renewcommand{\theequation}{\arabic{section}.\arabic{equation}}

\newcommand{\Q}{\widetilde{Q}}
\newcommand{\calP}{{\mathcal P}}
\newcommand{\calQ}{{\mathcal Q}}

\newcommand{\R}{{\mathbb R}}

\newcommand{\X}{\mathbf{X}}
\renewcommand{\P}{\mathbb{P}}

\newcommand{\PP}{\widetilde{P}}

\newcommand{\x}{\mathbf{x}}

\newcommand{\e}{{\mathrm e}}
\newcommand{\E}{{\mathbb E}}

\newcommand{\n}{\mathbf n}

\renewcommand{\P}{\mathbb P}
\newcommand{\p}{\widetilde{p}}

\newcommand{\trho}{\widetilde{\rho}}

\begin{document}

\title{Trapping of an active Brownian particle at a partially absorbing wall}
\author{ \em
P. C. Bressloff, \\ Department of Mathematics, 
University of Utah \\155 South 1400 East, Salt Lake City, UT 84112}

\maketitle
\begin{abstract} 

Active matter concerns the self-organization of energy consuming elements such as motile bacteria or self-propelled colloids. A canonical example is an active Brownian particle (ABP) that moves at constant speed while its direction of motion undergoes rotational diffusion. When ABPs are confined within a channel, they tend to accumulate at the channel walls, even when inter-particle interactions are ignored. Each particle pushes on the boundary until a tumble event reverses its direction. The wall thus acts as a sticky boundary. In this paper we consider a natural extension of sticky boundaries that allows for a particle to be permanently killed (absorbed) whilst attached to a wall. In particular, we investigate the first passage time (FPT) problem for an ABP in a two-dimensional channel where one of the walls is partially absorbing. Calculating the exact FPT statistics requires solving a non-trivial two-way diffusion boundary value problem (BVP). We follow a different approach by separating out the dynamics away from the absorbing wall from the dynamics of absorption and escape whilst attached to the wall. Using probabilistic methods, we derive an explicit expression for the MFPT of absorption, assuming that the arrival statistics of particles at the wall are known. Our method also allows us to incorporate a more general encounter-based model of absorption.

\end{abstract}

\section{Introduction}

A major topic of current interest within the general field of non-equilibrium systems is {\em active matter},  which is typically described in terms of a collection of elements that consume energy in order to move or to
exert mechanical forces \cite{Ramaswamy10,Roman12,Solon15,Bechinger16}. Examples include animal flocks or herds \cite{Attanasi14}, motility-based phase separation \cite{Cates15}, bacterial suspensions \cite{Dombrowski04,Kaiser14}, synthetically manufactured self-propelled colloids \cite{Naryan07,Palacci10,Bricard13}, and components of the cellular cytoskeleton \cite{Schaller11}. In many cases the individual particles have an intrinsic orientation and can exhibit long-range orientational interactions mediated by some sensing mechanism or by coupling hydrodynamically to the surrounding medium \cite{Vicsek12}. 

In order to gain theoretical insights into the behavior of active matter, it is often useful to consider  simplified models of the individual particles, in particular, either a run-and-tumble particle (RTP) or an active Brownian particles (ABP) \cite{Solon15}. These two models provide an analytically tractable framework for studying self-organizing phenomena such as the accumulation of active particles at walls, which can occur even if inter-particle interactions are ignored \cite{Bechinger16}. Let $\X(t)\in \R^2$ and $\Theta(t)\in [0,2\pi]$ denote the position and orientation of an active particle in two dimensions. In the case of an RTP, the dynamics is described by the stochastic equation
\begin{equation}
\label{rtp}
\frac{d\X}{dt} =v_0 \n(\Theta(t)),\quad \n(\theta)=(\cos \theta,\sin \theta),
\end{equation}
where $v_0$ is the speed of the particle and the orientation $\Theta(t)$ randomly switches between a finite set of states $\{ \theta_1,\ldots, \theta_n\}$ according to a Markov chain. Mathematically speaking, equation (\ref{rtp}) is an example of a velocity jump process. (In one dimension (1D), equation (\ref{rtp}) reduces to a two state velocity jump process, in which the particle switches between the velocity states $\pm v_0$.) Turning to a 2D model of an ABP, the dynamics evolves according to a stochastic differential equation (SDE) of the form \cite{Basu19}
\begin{equation}
d\X(t)=v_0 \n(\theta(t))+\sqrt{2\overline{D}}d\overline{\bf W}(t),\quad d\Theta(t) =\sqrt{2D}dW(t),
\label{abp}
\end{equation}
where $\overline{\bf W}(t)=(\overline{W}_x(t),\overline{W}_y(t))$. The stochastic variables $\overline{W}_x(t),\overline{W}_y(t),{W}(t)$ are independent Wiener processes, $\overline{D}$ is the translational diffusivity, and $D$ is the rotational diffusivity (with units of inverse time).
 
Both models exhibit accumulation at the boundaries of a confinement domain, even at the single particle level (see Ref. \cite{Angelani17} for an RTP in an interval and Refs. \cite{Lee13,Wagner17,Wagner19,Wagner22} for an ABP in a 2D channel.). This is due to the fact that whenever a particle hits a hard wall, it becomes stuck by pushing on the boundary until a tumble event reverses its direction. At the multi-particle level this results in a pressure being exerted on the confining walls. An equivalent way to formulate the accumulation process is in terms of a sticky boundary condition.  
That is, whenever the particle collides with a wall, it remains attached to the wall for a random time interval that is determined by the tumbling dynamics. If the escape time back into the bulk is zero then the boundary is totally reflecting, whereas if the particle never escapes then the boundary is totally absorbing. The intermediate case is known as a sticky boundary condition. Sticky boundary conditions 
also arise within the context of the growth and shrinkage of polymer filaments such as microtubules and actin-rich filopidia (cytonemes) in confined 1D domains \cite{Zelinski12,Mulder12,Bressloff19}. For example, a nucleation site for polymer formation can be modeled as a sticky boundary, as can the temporary attachment of a filament to a cell wall during cytoneme-based embryogenesis \cite{Bressloff19}. Denoting the position of the polymer tip at time $t$ by the variable $X(t)$, one can model the dynamics of $X(t)$ in terms of a two-state velocity jump process similar to a 1D RTP. 

In this paper we are interested in a natural extension of the standard sticky boundary condition that allows for a particle to be permanently killed (absorbed) whilst attached to a wall. In the case of a bacterium this could be due to some noxious substance within the wall, whereas in the case of a growing filament it could represent destruction of the nucleation site itself. (Note that the killing process is distinct from the temporary absorption of a particle at a sticky boundary.)  In contrast to the dynamics of an active particle in a bounded domain without killing, there no longer exists a non-trivial steady state density for particle position and orientation. Instead, quantities of interest include the mean first passage time (MFPT) for permanent absorption, and possibly higher order statistics. The FPT problem for killing of an RTP in 1D has been studied in Refs. \cite{Angelani17,Bressloff23}. The main goal of the current paper is to investigate the analogous FPT problem for an ABP in a 2D channel where one of the walls is partially absorbing. One of the standard methods for calculating the MFPT for an SDE such as equation (\ref{abp}) is to solve the corresponding Fokker-Planck equation for the particle probability density by Laplace transforming with respect to time $t$. However, in the case of an ABP confined to a 2D channel, the resulting boundary value problem (BVP) is non-trivial since one cannot appeal to standard Sturm-Liouville theory. More specifically, the BVP is an example of a so-called two-way diffusion problem \cite{Fisch80,Beals83,Beals85}, since the boundary conditions at the channel walls are only defined on the orientation half spaces $\theta \in {\mathcal I}_-:=(\pi/2,3\pi/2)$ and $\theta \in {\mathcal I}_+:=(-\pi/2,\pi/2)$. 

One way to proceed would be to adapt the hybrid analytical/numerical method recently developed to solve the two-way diffusion problem for the steady-state density in the absence of killing \cite{Wagner17,Wagner19,Wagner22}. Although this hybrid approach does generate solutions that are consistent with simulations of the full system in specific examples, there are a number of technical difficulties in establishing convergence of the solutions. These difficulties are multiplied when solving the time-dependent BVP in Laplace space. Therefore, we follow a different approach here by separating out the dynamics away from the absorbing wall from the dynamics of absorption and escape whilst attached to the wall. In particular, we adapt a probabilistic method for solving FPT problems that we previously developed for finite-state velocity jump processes in the presence of sticky boundaries \cite{Bressloff19}. Assuming that the arrival statistics of particles at the wall are known, we show how calculating the MFPT for permanent absorption reduces to solving a simpler angular FPT that determines the conditional MFPT of absorption and escape (return to the channel) whilst a particle is attached to the wall. The latter calculation is a non-trivial extension of a trapping model analyzed in Ref. \cite{Moen22}. One particular advantage of separating out the boundary and bulk dynamics is that it allows us to incorporate a more general model of killing.

The structure of the paper is as follows. In section 2 we write down the Fokker-Planck equation for an ABP
in a 2D channel and formulate the basic FPT problem. In section 3, we derive a formula for the MFPT using probabilistic methods. This expresses the MFPT in terms of the absorption and escape statistics of a particle attached to the wall. The latter are analyzed in section 4 where we also present some illustrative examples. In section 5, we extend the analysis to a more general model of absorption using an encounter-based method \cite{Grebenkov20,Grebenkov22,Bressloff22a,Bressloff22b,Bressloff22c}. Finally, in appendices A-C we outline how to extend the hybrid analytical/numerical method of Refs. \cite{Wagner17,Wagner19,Wagner22} to the non-stationary case, highlighting some of the technical difficulties.

\setcounter{equation}{0}
\section{Active Brownian particle confined to a semi-infinite channel}

\begin{figure}[t!]
\centering
\includegraphics[width=8cm]{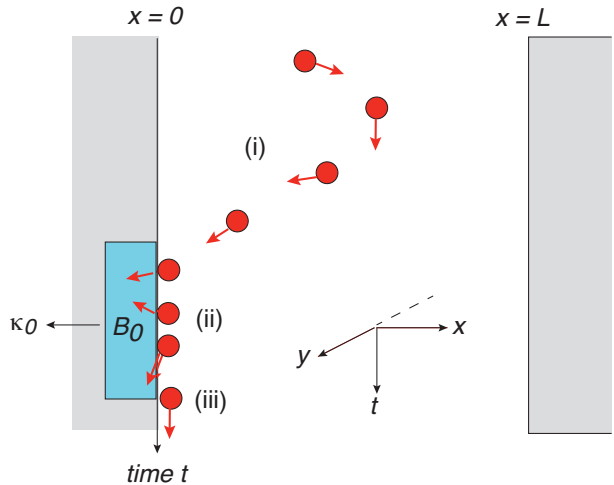}
\caption{Schematic representation of the time evolution of the position $x$ and the velocity direction (indicated by arrows) of an ABP confined to an infinitely long 2D channel of width $L$. (i) Motion within the bulk. (ii) The particle hits the wall at $x=0$ and remains stuck at the wall (in the bound state $B_0$) until rotational diffusion allows it to escape the wall (iii). Prior to escape, the particle may be absorbed at a rate $\kappa_0$. The right-hand wall is assumed to be non-absorbing, so that if the particle exists in the corresponding bound state $B_L$ then it always returns to the bulk.}
\label{fig1}
\end{figure}

Consider an ABP confined to a 2D channel $\Omega\subset \R^2$ of width $L$ in the $x$ direction and of infinite extension in the $y$ direction. Let $\X(t)\in \Omega$ and $\Theta(t) \in [0,2\pi]$ denote the position and orientation of the particle at time $t$. These stochastic variables are taken to evolve according to the SDE (\ref{abp}). For simplicity, we will neglect translational diffusion by setting $\overline{D}=0$. Let $p(x,y,\theta,t)$ denote the probability density for the triplet $(X(t),Y(t),\Theta(t))$. The density evolves according to the Fokker-Planck equation
\begin{equation}
 \frac{\partial p(\x,\theta,t)}{\partial t}=-v_0\n(\theta)\cdot \nabla p(\x,\theta,t)+D\frac{\partial^2p(\x,\theta,t)}{\partial \theta^2},\quad \x\in \Omega, \quad \theta \in [0,2\pi].
\end{equation}
Given the translation invariance in the $y$ direction, we assume that $p$ is independent of $y$ so that the Fokker-Planck equation reduces to the quasi-one dimensional form:
\begin{subequations}
\begin{equation}
 \frac{\partial p(x,\theta,t)}{\partial t}=-v_0\cos\theta \frac{\partial p(x,\theta,t)}{\partial x}+D\frac{\partial^2p(x,\theta,t)}{\partial \theta^2},\quad x\in (0,L),\quad \theta \in [0,2\pi].
\label{FP}
\end{equation}

The particle will hit the wall at $x=0$ if it is traveling to the left ($\cos \theta <0$), whereas it will hit the wall at $x=L$ if it is traveling to the right ($\cos \theta >0)$. As soon as it hits the wall its linear velocity drops to zero but its orientation will continue to diffuse. The particle remains stuck at the wall until the orientation crosses one of the vertical directions, after which it reenters the bulk domain. In contrast to previous studies of ABPs, however, we take the left-hand wall to be partially absorbing at a rate $\kappa_0$, see Fig. \ref{fig1}. This means that prior to reinserting the bulk domain from the left-hand wall, the particle may be permanently absorbed. Let $Q_0(\theta,t)$ denote the probability density that the particle is attached to the wall at $x=0$ and has orientation $\theta$ ($\cos \theta <0$). Then
\begin{equation}
 \frac{\partial Q_0(\theta, t)}{\partial t}=D\frac{\partial^2Q_0(\theta,t)}{\partial \theta^2} 
-v_0\cos \theta  p(0,\theta,t)  -\kappa_0Q_0(\theta,t) ,\ \theta \in {\mathcal I}_-:= (\pi/2,3\pi/2).
\label{q}
\end{equation}
Equation (\ref{q}) is 
supplemented by the absorbing boundary conditions given by $Q_0(\pm \pi/2,t)=0$, which signal the reinsertion of the particle into the bulk domain. The absorbing boundary conditions mean that the net flux from the left-hand wall back into the bulk is 
\begin{align}
\label{J}
 v_0 \cos \theta p(0,\theta,t)&= D\frac{\partial Q_0(\pi/2,t)}{\partial \theta}\delta(\theta-\pi/2+\epsilon)-D\frac{\partial Q_0(-\pi/2,t)}{\partial \theta}\delta(\theta+\pi/2-\epsilon), 
\end{align}
where $0<\epsilon \ll 1$. The small parameter $\epsilon$ is introduced to avoid the singularities at $\pm \theta =\pi/2$. However, the resulting solution is well defined in the limit $\epsilon \rightarrow 0$.
Similarly, the probability density $Q_L(\theta,t)$ that the particle is attached to the wall at $x=L$ and has orientation $\theta$ ($\cos \theta >0$) evolves according to the equation
\begin{equation}
\label{qL}
 \frac{\partial Q_L(\theta, t)}{\partial t}=D\frac{\partial^2Q_L(\theta,t)}{\partial \theta^2} 
+v_0\cos \theta  p(L,\theta,t) ,\ \theta \in {\mathcal I}_+:= (-\pi/2,\pi/2),
\end{equation}
with $Q_L(\pm \pi/2,t)=0$ and
\begin{align}
v_0 \cos \theta p(L,\theta,t)&=-D\frac{\partial Q_L(\pi/2,t)}{\partial \theta}\delta(\theta-\pi/2-\epsilon)\nonumber\\
&\quad +D\frac{\partial Q_L(-\pi/2,t)}{\partial \theta}\delta(\theta+\pi/2+\epsilon).
\label{JL}
\end{align}
\end{subequations}

A number of recent studies have analyzed the steady-state version of the above equations in the case of no absorption ($\kappa_0=0$) using the theory of two-way diffusion processes\cite{Lee13,Wagner17,Wagner19}. However, in the presence of absorption, the probability densities converge to zero in the large time limit. That is, there does not exist a non-trivial steady state. A fundamental quantity of interest is now the MFPT for absorption at $x=0$. Consider the survival probability 
\begin{equation}
\label{SP0}
S(t):=\int_0^{L}\int_{-\pi}^{\pi}p(x,\theta,t)d\theta dx+Q_0(t)+Q_L(t),
\end{equation}
with
\begin{equation}
Q_0(t):=\int_{{\mathcal I}_-}Q_0(\theta,t)d\theta,\quad Q_L(t):=\int_{{\mathcal I}_+}Q_L(\theta,t)d\theta.
\end{equation}
In particular $Q_0(t)$ and $Q_L(t)$ are the probabilities that at time $t$ the particle is bound to the wall at $x=0$ and $x=L$, respectively.
Differentiating both sides with respect to $t$ using equations (\ref{FP}), (\ref{q}) and (\ref{qL}), we have
\begin{align}
\frac{\partial S}{\partial t}&=\int_0^L\int_0^{2\pi}\left [-v_0\cos\theta \frac{\partial p(x,\theta,t)}{\partial x}+D\frac{\partial^2p(x,\theta,t)}{\partial \theta^2}\right ]d\theta dx\ \nonumber \\
&\quad +\int_{{\mathcal I}_-}\left [D\frac{\partial^2Q_0(\theta,t)}{\partial \theta^2} 
-v_0\cos \theta  p(0,\theta,t) -\kappa_0 Q_0(\theta,t)\right ]d\theta\nonumber \\
&\quad +\int_{{\mathcal I}_+}\left [D\frac{\partial^2Q_L(\theta,t)}{\partial \theta^2} 
+v_0\cos \theta  p(L,\theta,t) \right ]d\theta=-\kappa_0 Q_0(t).
\label{SQ}
\end{align}
The terms involving $v_0\cos \theta$ cancel, whereas the contributions from the rotational diffusion terms
vanish due to periodicity with respect to $\theta$. 
Equation (\ref{SQ}) implies that the rate at which the survival probability decreases in time is equal to the absorption flux  $J_{0}(t)=\kappa_0Q_{0}(t)$ at $x=0$. Let $T$ denote the FPT for absorption. The FPT density is $f(t):=- {\partial S(t)}/{\partial t}=J_0(t)$,
and the corresponding MFPT is
\begin{align}
 \tau:= \E[ T] &= \int_0^{\infty}t f(t)dt= -\int_0^{\infty}tJ_0(t)dt =-\lim_{s\rightarrow 0} \frac{\partial}{\partial s}\widetilde{J}_0(s),
 \label{MFPT}
\end{align}
where $\widetilde{J}_0(s)=\int_0^{\infty}\e^{-st}J_0(t)dt$. Note that $\tau$ also depends on the initial conditions. 

Equation (\ref{MFPT}) suggests that one method for calculating the MFPT is to determine $\widetilde{J}_0(s)=\kappa_0\widetilde{Q}_0(s)$ by solving equations (\ref{FP})--(\ref{JL}) in Laplace space, and then taking the small-$s$ limit. In the appendices we outline how this could be achieved by extending the hybrid analytical/numerical scheme for solving two-way diffusion processes in steady state. However, it is clear that this method is analytically and computationally non-trivial. Therefore, in this paper we follow a different approach by separating out the dynamics away from the absorbing wall from the absorption and escape process whilst in the bound state $B_0$. This is similar in spirit to a recent study of the trapping of ABPs and RTPs at a wall \cite{Moen22}. These authors assume that a particle hits the wall at some angle $\theta_0$ and determine the angular first passage time statistics for escaping the wall and returning to the bulk; the wall is taken to be non-absorbing. In the case of a partially absorbing wall, the problem is considerably more difficult due to the fact that the particle may have to undergo multiple rounds of sticking to the wall and escaping into the bulk before finally being absorbed. We proceed by adapting a probabilistic method for solving FPT problems for finite-state velocity jump processes in the presence of sticky boundaries, which uses some classical concepts from probability theory, namely, conditional expectations and the strong Markov property \cite{Bressloff19}.

\section{Calculation of the killing MFPT}

\subsection{Probabilistic decomposition of the FPT}
\begin{figure}[t!]
\begin{center}
\includegraphics[width=13cm]{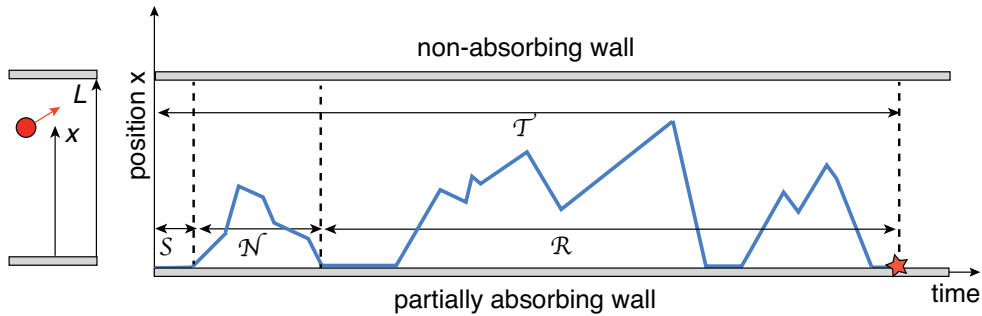}
\end{center}
\caption{Decomposition ${\mathcal T}={\mathcal S}+{\mathcal N}+{\mathcal R}$ of the FPT to be permanently absorbed or killed by the wall at $x=0$ starting at $x=0$ in the bound state $B_0$. Here ${\mathcal S}$ is the FPT to escape the wall, ${\mathcal N}$ is the FPT to return to the wall following an escape, and ${\mathcal R}$ is the FPT to be absorbed, given at least one escape event.  }
\label{fig2}
\end{figure}

The basic idea underlying the probabilistic method is to decompose a sample trajectory of an ABP into a sequence of events consisting of alternating rounds of binding to the wall at $x=0$ and subsequent excursions away from the wall into the domain $(0,L]$ until the particle is killed. An example trajectory $X(t)$ is shown in Fig. \ref{fig2}. For ease of notation, we take the particle to be in the bound state $B_0$ whenever $X(t)=0$, to be away from the wall whenever $X(t)>0$ and to be absorbed when $X(t)<0$. Consider the following set of conditional FPTs:
\begin{subequations}
\begin{align}
{\mathcal T}&=\inf\{t\geq 0; X(t)<0 \}, \\
{\mathcal S} &=\inf\{t\geq 0; X(t)>0\},\\
{\mathcal N}  &=\inf\{t\geq 0; X({\mathcal S}+t)=0\},\\
{\mathcal R}  &=\inf\{t\geq 0; X({\mathcal S}+{\mathcal N }+t)<0\}.
\end{align}
\end{subequations}
Introducing the set $\Omega=\{{\mathcal S}<{\mathcal T}\}$,
we can decompose the MFPT according to
\begin{align}
\tau:=\E[{\mathcal T}]&=\E[{\mathcal T}1_{\Omega^c}]+\E[{\mathcal T}1_{\Omega}] =\E[{\mathcal T}1_{\Omega^c}]+\E[({\mathcal S}+{\mathcal N} +{\mathcal R})1_{\Omega}],
\label{MFPT2}
\end{align}
where $\Omega^c$ is the complementary set of $\Omega$ and $1_{\Omega}$ denotes the indicator function, which ensures that expectation is only taken with respect to events that lie in $\Omega$. The first two terms $\E[{\mathcal T}1_{\Omega^c}]$ and $\E[{\mathcal S}1_{\Omega}]$ are completely determined by the tumbling dynamics at the left-hand wall. In particular, 
\begin{equation}
\E[{\mathcal T}1_{\Omega^c}]=\pi_{\rm abs} \tau_{\rm abs} \mbox{ and } \E[{\mathcal S}1_{\Omega}]={\pi}_{\rm esc}{\tau}_{\rm esc}.
\end{equation}
 Here $\tau_{\rm abs}$ and $\pi_{\rm abs}$ are the conditional FPT and splitting probability that the particle is absorbed before ever escaping the wall. Similarly, $\tau_{\rm esc}$ and $\pi_{\rm esc}$ are the conditional FPT and splitting probability that the particle escapes into the bulk domain before being absorbed. The third term is
\begin{equation}
\E[{\mathcal N}1_{\Omega}]=\E[{\mathcal N}|{\mathcal S}<{\mathcal T}]\P[\Omega]=\pi_{\rm esc}\overline{\tau},
\end{equation}
 where $\overline{\tau}$ is the MFPT for the particle to return to the left-hand wall. 
 
 In order to determine the final term $\E[{\mathcal R}1_{\Omega}]$, we assume that the particle starts at $x=0$ with an orientation
 $\theta_0\in {\mathcal I}_-$ that is distributed according to a density $g(\theta_0)$ that is identical to the distribution of 
 orientations of the particle when it returns to the wall. Under this simplification, we can now exploit an important property of the given stochastic process, namely, it satisfies the strong Markov property. Recall that a stochastic process $\{X(t)\}_{t\in T}$ is said to have the {Markov property} if the conditional probability distribution of future states of the process (conditional on both past and present states) depends only upon the present state, not on the sequence of events that preceded it. That is, for all $t'>t$ we have $\P[X_{t'}\leq x|X_{s},s\leq t]=\P[X_{t'}\leq x|X_{t}]$. 
 The {strong Markov property} is similar to the Markov property, except that the ``present'' is defined in terms of a first passage time (or more general stopping time). That is, given any finite-valued FPT ${\mathcal T}$, if the stochastic process $Y(t)=X(t+{\mathcal T})-X({\mathcal T}$ is independent of $\{X(s),s<{\mathcal T}\}$ and has the same distribution as $\widehat{Y}(t)=X(t)-X(0)$ then $X$ is said to satisfy the strong Markov property.  In terms of our current example, the strong Markov property implies that 
\begin{equation}
 \E[{\mathcal R}1_{\Omega}]=\E[{\mathcal R}|{\mathcal S}<{\mathcal T}]\P[\Omega]=\pi_{\rm esc}{\mathcal T}.
 \end{equation}
 That is, whenever the particle returns to the left-hand wall, it loses all memory of its previous excursions in the domain $(0,L]$. It follows that equation (\ref{MFPT2}) becomes
\begin{align}
\tau=\pi_{\rm abs} \tau_{\rm abs}+\pi_{\rm esc} \tau_{\rm esc}+\pi_{\rm esc}\overline{\tau}+\pi_{\rm esc}\tau,
\end{align}
which, after rearranging, yields the final result\footnote{Equation (\ref{MFPTf}) has an intuitive interpretation in terms of independent Bernouilli trials. Suppose that the particle makes $n$ excursions before being killed. This occurs with probability $\pi_{\rm abs}\pi_{\rm esc}^n$, and each excursion costs a mean amount of time $\tau_{\rm esc}+\overline{\tau}$. Hence, $\tau=\pi_{\rm abs} \tau_{\rm abs}+\pi_{\rm abs}\sum_{n=0}^{\infty}n(\tau_{\rm esc}+\overline{\tau})\pi_{\rm esc}^n$. Summing the geometric series recovers equation (\ref{MFPTf}). The advantage of the more rigorous derivation is that it can be generalized to more complicated scenarios involving multiple sticky boundaries in higher dimesnions, for example.
}
\begin{align}
\tau=\frac{\pi_{\rm abs} \tau_{\rm abs}+\pi_{\rm esc}( \tau_{\rm esc}+\overline{\tau})}{1-\pi_{\rm esc}}.
\label{MFPTf}
\end{align}
Note that if $\pi_{\rm esc}=0$ then $\pi_{\rm abs}=1$ and $\tau =\tau_{\rm abs}$. This means that the stochastic dynamics in $(0,L]$ does not play a role other than determining the initial density $g(\theta_0)$. On the other hand, if $\pi_{\rm esc}=1$ then the particle is never absorbed and $\tau=\infty$. Finally, if $\overline{\tau}=0$ (instant return), then the boundaries $\theta=\pm\pi/2$ are effectively reflecting and $\tau=1/\kappa_0$.

\subsection{Splitting probabilities and MFPTs for  absorption and escape at $x=0$.}

\begin{figure}[t!]
\centering
\includegraphics[width=8cm]{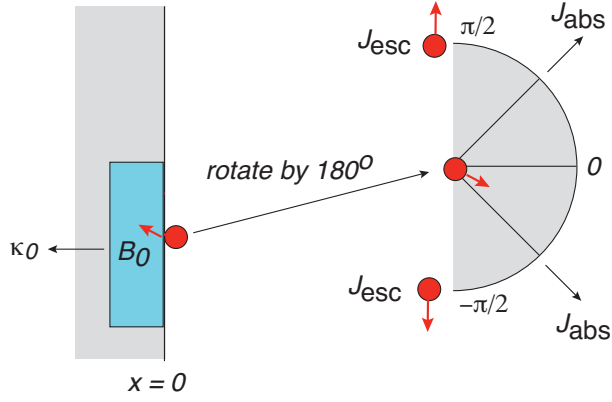}
\caption{Illustration of the angular FPT problem at the partially absorbing wall (after shifting $\theta \rightarrow \theta+\pi$). The particle starts at an orientation $\theta_0\in (-\pi/2,\pi/2)$ and undergoes rotational diffusion until either crossing the angles $\pm\pi/2$ and escaping into the bulk domain, or it is absorbed at a rate $\kappa_0$. The escape and absorption fluxes are also indicated.}
\label{fig3}
\end{figure}

So far we have simplified the FPT problem for absorption by assuming that the MFPT $\overline{\tau}$ for first return to $x=0$ and the distribution $g(\theta)$ of the orientation of a returning particle are known. The calculation of $\tau$ thus reduces to the problem of determining the statistics of absorption and escape when the particle is in the bound state and undergoing rotational diffusion. We proceed by generalizing the analysis for a non-absorbing wall presented in Moen {\em et al.}\cite{Moen22}. For convenience, we shift the orientation angle by $\pi$ so that the particle approaches the wall from the left and is stuck to the wall if $\theta \in (-\pi/2,\pi/2)$, see Fig. \ref{fig3}. Let $\rho(\theta,t|\theta_0)$ be the probability density that the bound particle has orientation $\theta \in (-\pi/2,\pi/2)$ at time $t$, given that it started with the orientation $\theta_0\in (-\pi/2,\pi/2)$. The corresponding Fokker-Planck equation is
\begin{equation}
\label{aFP}
\frac{\partial \rho}{\partial t}=D\frac{\partial^2\rho}{\partial \theta^2}-\kappa_0\rho,\quad \theta \in (-\pi/2,\pi/2);\quad \rho(\pm \pi/2,t|\theta_0)=0.
\end{equation}
Introducing the associated survival probability 
\begin{equation}
\label{Sang}
S(\theta_0,t)=\int_{-\pi/2}^{\pi/2}p(\theta,t|\theta_0)d\theta,
\end{equation} 
we have
\begin{align}
\frac{\partial S}{\partial t}&=\int_{-\pi/2}^{\pi/2}\frac{\partial \rho(\theta,t|\theta_0)}{\partial t}d\theta
=D\int_{-\pi/2}^{\pi/2}\frac{\partial^2 \rho(\theta,t|\theta_0)}{\partial^2 \theta}d\theta-\kappa_0S(\theta_0,t)\nonumber \\
&=D \left .\frac{\partial \rho(\theta,t|\theta_0)}{\partial \theta}\right |_{\theta=-\pi/2}-D \left .\frac{\partial \rho(\theta,t|\theta_0)}{\partial \theta}\right |_{\theta=\pi/2}-\kappa_0S(\theta_0,t).
\end{align}
We thus define the absorption and escape fluxes according to
\begin{equation}
\label{escape0}
J_{\rm esc}(\theta_0,t):=D \left .\frac{\partial \rho(\theta,t|\theta_0)}{\partial \theta}\right |_{\theta=\pi/2}-D \left .\frac{\partial \rho(\theta,t|\theta_0)}{\partial \theta}\right |_{\theta=-\pi/2},\,
J_{\rm abs}(\theta_0,t):=\kappa_0S(\theta_0,t).
\end{equation}
From these definitions, the splitting probabilities are
\begin{align}
\pi_{\rm abs}(\theta_0)&=\int_0^{\infty} J_{\rm abs}(\theta_0,t)dt=\widetilde{J}_{\rm abs}(\theta_0,0),\nonumber\\
\quad \pi_{\rm esc}(\theta_0)&=\int_0^{\infty} J_{\rm esc}(\theta_0,t)dt=1-\pi_{\rm abs}(\theta_0).
\label{split}
\end{align}
The conditional MFPTs for absorption satisfies the equation
\begin{equation}
\label{MFPTabs}
\pi_{\rm abs}(\theta_0)\tau_{\rm abs}(\theta_0)=\int_0^{\infty}tJ_{\rm abs}(\theta_0,t)dt
=-\lim_{s\rightarrow 0}\frac{\partial}{\partial s}\widetilde{J}_{\rm abs}(\theta_0,s).
\end{equation}
Similarly,
\begin{align}
\pi_{\rm esc}(\theta_0)\tau_{\rm esc}(\theta_0)&=\int_0^{\infty}tJ_{\rm esc}(\theta_0,t)dt
=-\int_0^{\infty}t\left (\frac{\partial S(\theta_0,t)}{\partial t}+\kappa_0S(\theta_0,t)\right ]dt\nonumber \\
&=\widetilde{S}(\theta_0,s)+\lim_{s\rightarrow 0}\frac{\partial}{\partial s}\widetilde{J}_{\rm abs}(\theta_0,s).
\label{MFPTesc}
\end{align}

\begin{figure}[b!]
\centering
\includegraphics[width=9cm]{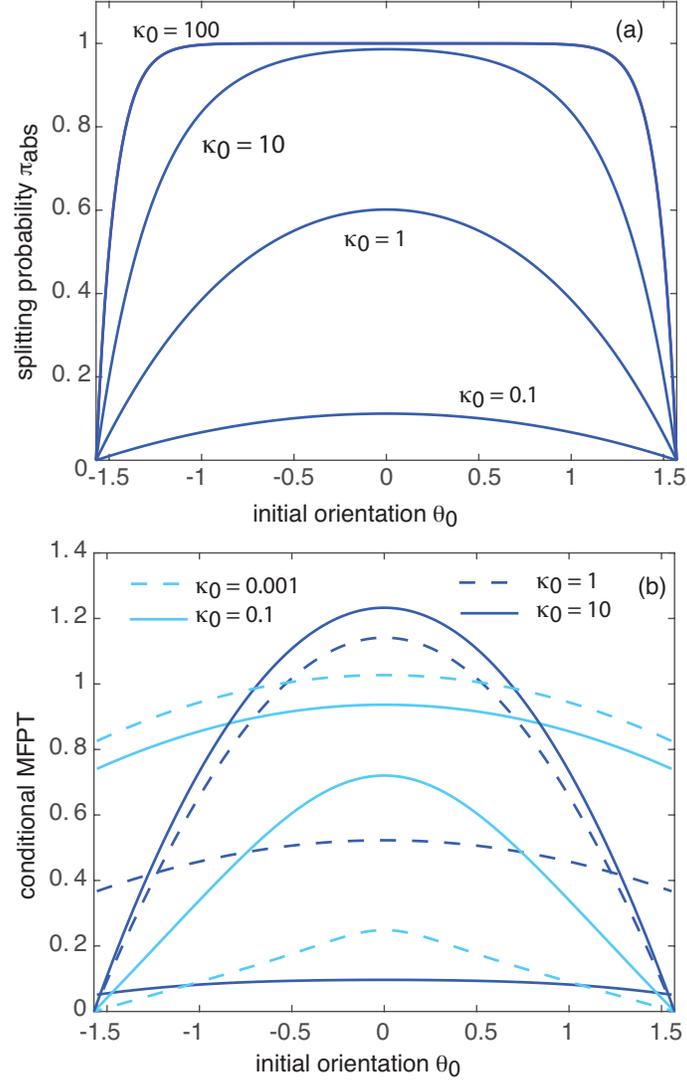}
\caption{Splitting probabilities and conditional MFPTs for the angular FPT problem. (a) Plots of $\pi_{\rm abs}(\theta_0)$ in equation (\ref{split}) as a function of the initial orientation for different killing rates $\kappa_0$. (b) Corresponding plots of $\tau_{\rm abs}(\theta_0)$ (flat curves) and $\tau_{\rm esc}(\theta_0)$ (steep curves), which are given by equations (\ref{MFPTabs}) and (\ref{MFPTesc}), respectively. The rotational diffusivity is $D=1$.}
\label{fig4}
\end{figure}

Laplace transforming equation (\ref{aFP}), we have
\begin{equation}
\label{aFPLT}
D\frac{\partial^2\trho}{\partial \theta^2}-(s+\kappa_0)\trho=-\delta(\theta -\theta_0),\quad \theta \in (-\pi/2,\pi/2); \quad \trho(\pm \pi/2,s|\theta_0)=0.
\end{equation}
It follows that $\trho(\theta,s|\theta_0)$ is the Dirichlet Green's function of the 1D modified Helmholtz equation. In particular,
\begin{equation}
\trho(\theta,s|\theta_0)=\left \{ \begin{array}{ll} \Xi \sinh(\gamma (\theta+\pi/2))\sinh(\gamma [\pi/2-\theta_0]), & -\pi/2 < \theta <\theta_0,\\
\Xi  \sinh(\gamma [\pi/2-\theta])\sinh(\gamma (\theta_0+\pi/2)), &  \theta_0 <\theta < \pi/2, \end{array}\right . , 
\label{rhop}\end{equation}
with 
\begin{equation}
\gamma=\sqrt{(s+\kappa_0)/D},\quad \Xi =\frac{1}{\sqrt{(s+\kappa_0)D}\sinh (\pi \gamma)}.
\end{equation}
The factor $\Xi$ is obtained from the flux discontinuity condition $\partial_{\theta}\trho(\theta,s|\theta_0)|_{\theta=\theta_0^-}-\partial_{\theta}\trho(\theta,s|\theta_0)|_{\theta=\theta_0^+}=1/D$ .
Hence, after some algebra, we find that
\begin{align}
\widetilde{S}(\theta_0,s)&=\int_{-\pi/2}^{\pi/2} \trho(\theta,s|\theta_0)d\theta =\frac{1}{ s+\kappa_0 }\left [1-\frac{\sinh(\gamma [\pi/2-\theta_0])+\sinh(\gamma [\pi/2+\theta_0])}{\sinh(\pi \gamma)}\right ],
\label{tilS}
\end{align}
and
\begin{align}
\frac{\partial \widetilde{S}(\theta_0,s)}{\partial s}&=- \frac{\widetilde{S}(\theta_0,s)}{s+\kappa_0}-\frac{[\pi/2-\theta_0]\cosh(\gamma [\pi/2-\theta_0])+(\theta_0+\pi/2)\cosh(\gamma [\theta_0+\pi/2])}{2D(s+\gamma)\sinh(\pi \gamma)}\nonumber \\
&\quad +\pi \cosh(\pi \gamma)\frac{\sinh(\gamma [\pi/2-\theta_0])+\sinh(\gamma [\pi/2+\theta_0])}{2D(s+\gamma)\sinh^2(\pi \gamma)}.
\label{tildS}
\end{align}
The splitting probabilities and conditional MFPTs are now obtained by setting $\widetilde{J}_{\rm abs}(\theta_0,s)=\kappa_0\widetilde{S}(\theta,s)$, and then substituting equations (\ref{tilS}) and (\ref{tildS}) into equations (\ref{split}), (\ref{MFPTabs}) and (\ref{MFPTesc}).
One can also check that if $s+\kappa_0\ll D$ then 
\begin{equation}
\widetilde{S}(\theta_0,s)=\frac{1}{2D}\left [\left (\frac{\pi}{2}\right )^2-\theta_0^2\right ]+O(\kappa_0+s).
\end{equation}
This means that $\pi_{\rm abs}\rightarrow 0$ in the limit $\kappa_0\rightarrow 0$ as expected. Moreover, the classical result $\tau_{\rm esc} =[(\pi/2)^2-\theta_0^2)]/2D$ is recovered when $\kappa_0=0$.

\subsection{Results}

In Fig. \ref{fig4} we plot the splitting probability $\pi_{\rm abs}(\theta_0)$ and the conditional MFPTs $\tau_{\rm abs}(\theta_0)$ and $\tau_{\rm esc}(\theta_0)$ as a function of the initial orientation $\theta_0$ for various killing rates $\kappa_0$. These plots are based on equations (\ref{split}), (\ref{MFPTabs}) and (\ref{MFPTesc}), respectively. Fig. \ref{fig4}(a) shows that $\pi_{\rm abs}(\theta_0)\rightarrow 0$ as $\theta_0\rightarrow \pm \pi/2$, consistent with the fact that the particle immediately escapes. As expected, $\pi_{\rm abs}(\theta_0)$ is a unimodal function of $\theta_0$ with a maximum at $\theta_0=0$, where the particle starts at an orientation directed perpendicularly into the wall. In addition, increasing $\kappa_0$ sharpens the curve so that $\pi_{\rm abs}(\theta_0)\approx 1$ away from the orientations $\theta_0=\pm \pi/2$. The conditional MFPT $\tau_{\rm esc}(\theta_0)$ is also a unimodal function of $\theta_0$ with a maximum at $\theta_0=0$ and vanishing at $\theta_0=\pm\pi/2$, see Fig. \ref{fig4}(b). Increasing $\kappa_0$ sharpens the curve and increases the value at the maximum. On the other hand, the conditional MFPT $\tau_{\rm abs}(\theta_0)$  has a much weaker dependence on $\theta_0$ and is a decreasing function of $\kappa_0$. Note that the non-zero values of $\lim_{\theta_0=\pm \pi/2}\tau_{\rm abs}(\theta_0)$ should be interpreted as left and right limits respectively.

\begin{figure}[t!]
\centering
\includegraphics[width=9cm]{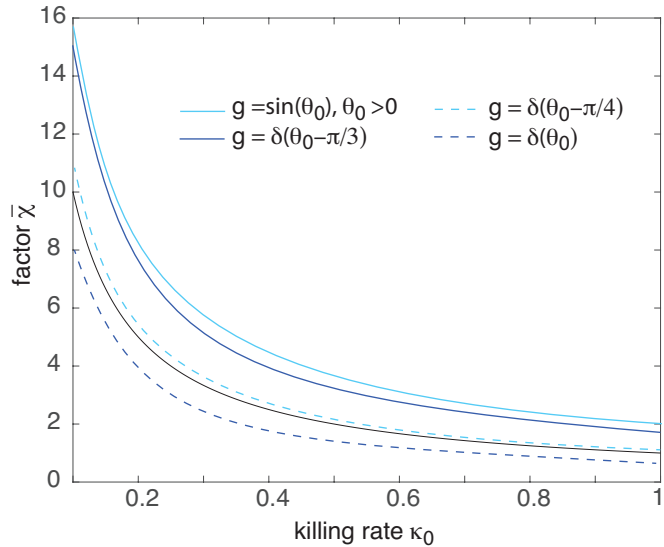}
\caption{Coefficient $\chi$ determining the effects of multiple excursions away from the wall on the MFPT $\tau$ of equation (\ref{MFPTff}). The thin solid curve is a plot of the function $1/\kappa_0$. The rotational diffusivity is $D=1$.}
\label{fig5}
\end{figure}

Now suppose that we substitute equations (\ref{split}), (\ref{MFPTabs}) and (\ref{MFPTesc}) into the formula (\ref{MFPTf}) for the MFPT $\tau$ of the full absorption problem, given an initial distribution $g(x_0)$:
\begin{align}
\tau=\frac{1}{\kappa_0}+\frac{[1-\kappa_0\overline{S}]\overline{\tau}}{\kappa_0\overline{S}},\quad \overline{S}:=\int_{-\pi/2}^{\pi/2}g(\theta_0)\widetilde{S}(\theta_0,0)d\theta_0.
\label{MFPTff}
\end{align} 
The first contribution to $\tau$ is the expected time for absorption if the particle never escapes the wall, whereas the second terms sums over all possible excursions into the domain $(0,L]$. In Fig. \ref{fig5} we plot the factor $\overline{\chi}:=(1-\kappa_0\overline{S})/(\kappa_0\overline{S})$ multiplying the mean excursion time $\overline{\tau}$ for several choices of $g(\theta_0)$. Clearly $\overline{\chi}$ is a decreasing function of $\kappa_0$, since the likelihood of an excursion away from the wall is reduced. If $g(\theta_0)=\delta(\theta_0-\phi)$ then the contribution from excursions is smallest when $\phi=0$ and increases as $\phi \rightarrow \pm  \pi/2$. We compare the Dirac delta distribution with a smooth example $g(\theta_0)=\sin(\theta_0)H(\theta_0)$, where $H$ is the Heaviside function.

\section{Generalized absorption scheme}

Another advantage of the decomposition performed in section 3 is that it is relatively straightforward to incorporate a more general absorption mechanism at the wall using an encounter-based method \cite{Grebenkov20,Grebenkov22,Bressloff22a,Bressloff22b,Bressloff22c}. (Similarly, the stochastic tumbling of the bound ABP could be modified by taking a different rotational diffusivity than when away from the wall, for example.) Following along similar lines to Refs. \cite{Bressloff22a,Bressloff22b}, we model the angular diffusion process in terms of a generalized propagator, which is defined to be the joint probability density for particle orientation $\Theta(t)$ and the so-called occupation time $A(t)$ in the absence of absorption at the wall. The occupation time is a Brownian functional \cite{Ito65,Majumdar05} defined according to 
\begin{equation}
\label{occ}
A(t)=\int_{0}^tI_{[-\pi/2,\pi/2]}(\Theta(\tau))d\tau .
\end{equation}
Here $I_{[-\pi/2,\pi/2]}(\theta)$ denotes the indicator function of the set $[-\pi/2,\pi/2]$, that is, $I_{[-\pi/2,\pi/2]}(\theta)=1$ if $\theta\in [-\pi/2,\pi/2]$ and is zero otherwise. Hence, $A(t)$ specifies the amount of time the particle spends within $[-\pi/2,\pi/2]$ over the time interval $[0,t]$. We also take $\Theta(0)=\theta_0$ and $A(0)=0$.
Denoting the generalized propagator by $P(\theta,a,t|\theta_0)$ it can be shown that \cite{Bressloff22a}
\begin{subequations}
\begin{align}
\label{Pocca}
 &\frac{\partial P(\theta,a,t|x_0)}{\partial t}=D\frac{\partial^2 P(\theta,a,t|\theta_0)}{\partial \theta^2} -\left (\frac{\partial P}{\partial a}(\theta,a,t|\theta_0) +\delta(a)P(\theta,0,t|\theta_0) \right ) \\
 &\mbox{ for } -\pi/2<\theta<\pi/2 \mbox{ and }P(\pm \pi/2,a,t|\theta_0)=0.
\label{Poccc}
\end{align}
\end{subequations}
Next we introduce the killing time
\begin{equation}
\label{TA}
\widehat{\mathcal T}=\inf\{t>0:\ A(t) >\widehat{A}\},
\end{equation}
where $\widehat{A}$ is a random variable with $\P[\widehat{A}>a]=\Psi(a)$. Note that $\widehat{\mathcal T}=\infty$ if the particle escapes from the wall before being killed. It can be shown that the marginal probability density for particle orientation $\Theta(t) $ is \cite{Bressloff22a}
\begin{subequations}
\begin{align}
\label{peep}
\rho^{\Psi}(\theta,t|\theta_0)&=\int_0^{\infty}\Psi(a) P(\theta,a,t|\theta_0)da,\ -\pi/2\leq \theta \leq \pi/2.
\end{align}
\end{subequations}

Given the marginal probability density, we introduce the survival probability $S^{\Psi}(\theta_0,t)$ that the particle hasn't either escaped or been been killed up to time $t$, given that it started at $\theta_0$:
\begin{align}
\label{Socc}
 S^{\Psi}(\theta_0,t)&= \int_{-\pi/2}^{\pi/2} \rho^{\Psi}(\theta,t|\theta_0)d\theta =\int_0^{\infty}\Psi(a) \int_{-\pi/2}^{\pi/2} P(\theta ,a,t|\theta_0)d\theta\, da .
\end{align}
Differentiating both sides with respect to $t$, and using equations (\ref{Pocca})--(\ref{Poccc}) implies that
\begin{align}
 &\frac{\partial S^{\Psi}(x_0,t)}{\partial t}=\int_0^{\infty}\Psi(a) \int_{-\pi/2}^{\pi/2}D\frac{\partial^2 P(\theta,a,t|\theta_0)}{\partial \theta^2}d\theta\, da\nonumber \\
 &\hspace{3cm}- \int_0^{\infty} \Psi(a)\int_{-\pi/2}^{\pi/2}\left [\frac{\partial P}{\partial a}(\theta,a,t|\theta_0) +\delta(a)P(\theta,0,t|\theta_0)\right ]d\theta\, da\nonumber \\
 &=\left . D\frac{\partial P(\theta,a,t|\theta_0)}{\partial \theta}\right |_{\theta=\pi/2}-\left . D\frac{\partial P(\theta,a,t|\theta_0)}{\partial \theta}\right |_{\theta=-\pi/2}\nonumber \\
&\quad  - \int_0^{\infty} \psi(a) \int_{-\pi/2}^{\pi/2}Q(x,a,t|x_0)dx\, da .
\label{Qocc}
\end{align}
We have used integration by parts and set $\psi(a)=-\Psi(a)$. The escape and absorption currents of equation (\ref{escape0}) become
\begin{subequations}
\label{escape}
\begin{align}
J^{\Psi}_{\rm esc}(\theta_0,t)&=D\left . \frac{\partial P(\theta,a,t|\theta_0)}{\partial \theta}\right |_{\theta=\pi/2}-\left . D\frac{\partial P(\theta,a,t|\theta_0)}{\partial \theta}\right |_{\theta=-\pi/2} ,\\ 
J^{\Psi}_{\rm abs}(\theta_0,t)&=\int_0^{\infty} \psi(a) \int_{-\pi/2}^{\pi/2}P(\theta,a,t|\theta_0)d\theta\, da .
\end{align}
\end{subequations}

The crucial step in the encounter-based method is to note that in the case of an exponential distribution $\Psi(a)=\e^{-za}$, the BVP for the marginal density is identical to equation (\ref{aFP}) with $z$ playing the role of the absorption rate $\kappa_0$. On the other hand, equation (\ref{peep}) shows that using an exponential distribution is equivalent to Laplace transforming the propagator with respect to the occupation time variable $a$. If we also Laplace transform with respect to time $t$ and set
\begin{subequations}
\begin{align}
\widetilde{\widetilde{P}}(\theta,z,s|\theta_0)&=\int_0^{\infty} \e^{-za}\left [\int_0^{\infty} \e^{-st}P(\theta,a,t|\theta_0)dt\right ]da,
\end{align}
\end{subequations}
then we can make the identification $\widetilde{\widetilde{P}}(\theta,z,s|\theta_0)=\trho(\theta,z,s|\theta_0)$, where
$\trho(\theta,z,s|\theta_0)$ is the Green's function (\ref{rhop}) for $\kappa_0=z$. It follows that
\begin{align}
\label{Jesc2}
\widetilde{J}^{\Psi}_{\rm esc}(\theta_0,s)&= D\left .   \frac{\partial[\PP(\theta,a,s|\theta_0)}{\partial \theta}\right |_{\theta=\pi/2}-\left . D \frac{\partial \PP(\theta,a,s|\theta_0)}{\partial \theta}\right |_{\theta=-\pi/2},\\
\label{Jabs2}
\widetilde{J}^{\Psi}_{\rm abs}(\theta_0,s)&=\int_0^{\infty} \psi(a) \left [\int_{-\pi/2}^{\pi/2}\PP(\theta,a,s|\theta_0)d\theta \right ]da,\\
\label{S2}
\widetilde{S}^{\Psi}(\theta_0,s)&=\int_0^{\infty} \Psi(a) \left [\int_{-\pi/2}^{\pi/2}\PP(\theta,a,s|\theta_0)d\theta \right ]da,
\end{align}
where
$\PP(\theta,a,s|\theta_0)= {\mathcal L}_a^{-1}\trho(\theta,z,s|\theta_0)$,
and ${\mathcal L}_a^{-1}$ denotes the inverse Laplace transform.
 (We are assuming that integration and differentiation commute with the inverse Laplace operator.) 
 The splitting probabilities and conditional MFPTs are then obtained along analogous lines to the derivation of equations (\ref{split}), (\ref{MFPTabs}) and (\ref{MFPTesc}):
\begin{align}
\pi^{\Psi}_{\rm abs}(\theta_0)&=\widetilde{J}^{\Psi}_{\rm abs}(\theta_0,0),
\quad \pi_{\rm esc}^{\Psi}(\theta_0)=1-\pi_{\rm abs}^{\Psi}(\theta_0),
\label{ssplit}\\
\label{sMFPTabs}
\pi^{\Psi}_{\rm abs}(\theta_0)\tau^{\Psi}_{\rm abs}(\theta_0)&
=-\lim_{s\rightarrow 0}\frac{\partial}{\partial s}\widetilde{J}_{\rm abs}^{\Psi}(\theta_0,s),\\
\pi^{\Psi}_{\rm esc}(\theta_0)\tau^{\Psi}_{\rm esc}(\theta_0)&=\widetilde{S}^{\Psi}(\theta_0,s)+\lim_{s\rightarrow 0}\frac{\partial}{\partial s}\widetilde{J}^{\Psi}_{\rm abs}(\theta_0,s).
\label{sMFPTesc}
\end{align}

 For the sake of illustration, suppose that $\theta_0=0$. The MFPT of equation (\ref{MFPTf}) becomes
 \begin{align}
\tau^{\Psi}=\frac{\widetilde{S}^{\Psi}(0,0)}{\widetilde{J}^{\Psi}(0,0)}+\left [\frac{1-\widetilde{J}^{\Psi}(0,0)}{\widetilde{J}^{\Psi}(0,0)}\right ]\overline{\tau}.
\label{MFPTen}
\end{align}
From equation (\ref{tilS}) we have
\begin{align}
\int_{-\pi/2}^{\pi/2}\trho^{\Psi}(\theta,z,s|0) d\theta &=\frac{1}{ s+z }\left [1-\frac{1}{\cosh(\pi \gamma(z)/2)}\right ],\quad \gamma(z)=\sqrt{(s+z)/D}.
\end{align}
Expressing the inverse Laplace transform in terms of the Bromwich integral yields
\begin{align}
 \int_{-\pi/2}^{\pi/2}\PP(\theta,a,s|0)&=\frac{1}{2\pi i  }\int_{c-i\infty}^{c+i\infty} \e^{za}\frac{1}{ s+z }\left [1-\frac{1}{\cosh(\pi \gamma(z)/2)}\right ] dz,
 \label{brom}
\end{align}
with $c$, $c>0$, chosen so that the Bromwich contour is to the right of all singularities. 
The Bromwich integral (\ref{brom}) can be evaluated by closing the contour in the complex $z$-plane and using the Cauchy residue theorem. Since $\cosh(\pi \gamma(z)/2)$ is an even function of $\gamma(z)$, it follows that $z=-s$ is not a branch point and $\trho(\theta,z,s|\theta_0)$ is single-valued. Moreover, the singularity at $z=-s$ is removable. The resulting contour thus encloses a countably infinite number of poles, which correspond to the zeros of the function $\cosh(\pi \gamma(z)/2)$:
\begin{equation}
\sqrt{s+z}=(2m-1) i\sqrt{D}\implies z= -s-(2m-1)^2D,\quad m\geq 1.
\end{equation}
Applying Cauchy's residue theorem, we find that
\begin{equation}
 \int_{-\pi/2}^{\pi/2}\PP(\theta,a,s|0)d\theta =\frac{4}{\pi }\sum_{n\geq 1} \frac{(-1)^{n+1}}{2n-1}\e^{-sa}\e^{-(2n-1)^2 Da}.
\end{equation}
Finally, substituting into equations (\ref{Jabs2}) and (\ref{S2}) gives
\begin{align}
\label{resJ}
\widetilde{J}^{\Psi}_{\rm abs}(0,s)&=\frac{4}{\pi }\sum_{n\geq 1} \frac{(-1)^{n+1}}{2n-1}\widetilde{\psi}(s+(2n-1)^2 D),\\
\label{resS}
\widetilde{S}^{\Psi}(0,s)&=\frac{4}{\pi }\sum_{n\geq 1} \frac{(-1)^{n+1}}{2n-1}\widetilde{\Psi}(s+(2n-1)^2 D).\end{align}
We are assuming that the order of summation and integration can be reversed.

\begin{figure}[b!]
\centering
\includegraphics[width=9cm]{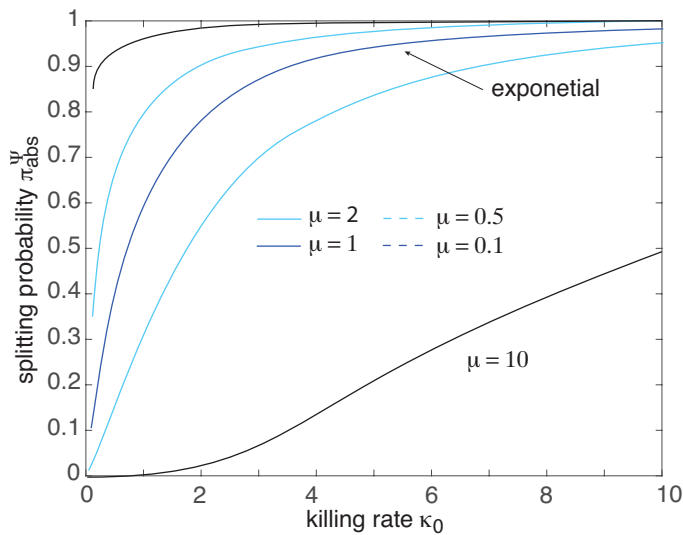}
\caption{Plot of the generalized splitting probability $\pi_{\rm abs}^{\Psi}(0)=\widetilde{J}^{\Psi}_{\rm abs}(0,0)$ as a function of the parameter $\kappa_0$ for various values of $\mu$. The Markovian (exponential) case is recovered when $\mu =1$.}
\label{fig6}
\end{figure}

\begin{figure}[t!]
\centering
\includegraphics[width=9cm]{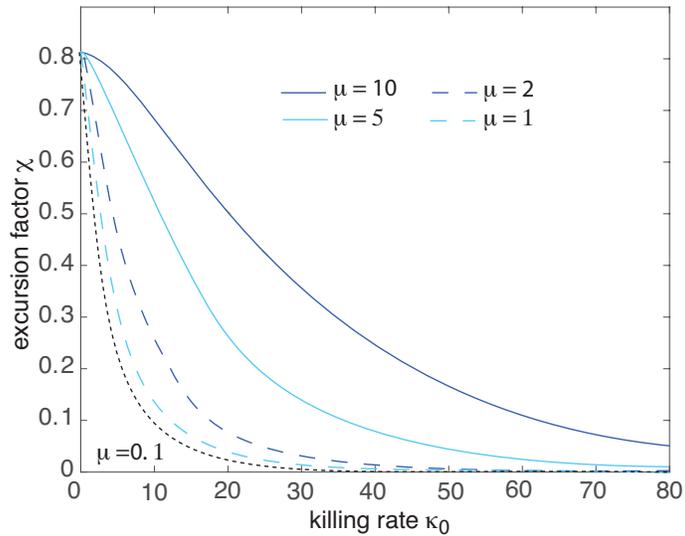}
\caption{Relative contribution  $\chi_0$ to the MFPT $\tau^{\Psi}$ of equation (\ref{MFPTen}) due to events without any excursions in the case of a gamma distribution. The term $\chi_0$  is plotted as a function of $\kappa_0$ for various values of $\mu$. In the exponential case ($\mu=1$), we have $\chi_0=1/\kappa_0$.}
\label{fig7}
\end{figure}

In order to illustrate the generalized absorption model, suppose that $\psi(a)$ is given by the gamma distribution:
\begin{equation}
\label{psigam}
\psi(a)=\frac{\kappa_0(\kappa_0a)^{\mu-1}\e^{-\kappa_0 a}}{\Gamma(\mu)},\  \mu >0,
\end{equation}
where $\Gamma(\mu)$ is the gamma function. The corresponding Laplace transforms are
\begin{equation}
\widetilde{\psi} (z)=\left (\frac{\kappa_0}{\kappa_0+z}\right )^{\mu},\quad \widetilde{\Psi}(z)=\frac{1-\widetilde{\psi}(z)}{z}.
\end{equation}
Note that if $\mu=1$ then $\psi$ reduces to the exponential distribution with constant reactivity $\kappa_0 $. The parameter $\mu$ thus characterizes the deviation of $\psi(\ell)$ from the exponential case. If $\mu <1$ ($\mu>1$) then $\psi(\ell)$ decreases more rapidly (slowly) as a function of the local time $\ell$. The effective reaction time is given by the mean $\E[\ell]=\mu/\kappa_0$. In Fig. \ref{fig6} we plot the generalized splitting probability $\pi_{\rm abs}^{\Psi}(0)=\widetilde{J}^{\Psi}_{\rm abs}(0,0)$ as a function of the parameter $\kappa_0$ for various values of $\mu$. It can be seen that decreasing $\mu$ significantly increases $\pi_{\rm abs}^{\Psi}(0)$, which is consistent with the observation that the wall becomes more effective at killing the particle. This effect is clearly nonlinear in the sense that changing $\mu$ does not simply rescale the effective killing rate according to $\kappa_0\rightarrow \kappa_0/\mu$. 
In Fig. \ref{fig7} we plot
\begin{equation}
\chi:=\frac{\tau^{\Psi}|_{\overline{\tau}=1}-\tau^{\Psi}|_{\overline{\tau}=0}}{\tau^{\Psi}|_{\overline{\tau}=0}}=\frac{1-\widetilde{J}^{\Psi}(0,0)}{\widetilde{S}^{\Psi}(0,0)},
\end{equation} 
as a function of $\kappa_0$ for different values of $\mu$. The factor $\chi$ quantifies the relative contribution of  excursions to the MFPT $\tau^{\Psi}$ in equation (\ref{MFPTen}).  That is, for a mean excursion time $\overline{\tau}$ we have $\tau^{\Psi}=\tau^{\Psi}|_{\overline{\tau}=0}[1+\chi \overline{\tau}]$. Again, changing $\mu$ fo fixed $\kappa_0$ has a nonlinear effect on the MFPT.

\section{Conclusion}  

In this paper we developed a probabilistic method for studying the killing or permanent absorption of an ABP in a 2D channel with a partially absorbing wall. We proceeded by separating out the dynamics away from the absorbing wall from the absorption and escape process whilst the particle is attached to the wall.  This generated a sequence of events consisting of alternating rounds of binding to the wall and subsequent excursions into the bulk domain until the particle is killed. The calculation of the MFPT thus reduced to solving an effective first passage time problem on a finite interval, assuming that the mean excursion time $\overline{\tau}$ and the orientation distribution $g(\theta)$ of particles hitting the wall are known. The analysis was also extended to a more general, non-Markovian form of absorption. 

One possible limitation of our approach is that it requires specifying $\overline{\tau}$ and $g(\theta)$.
However, such quantities could be determined numerically for a given channel configuration, and then applied to a wide range of different models of the absorbing wall. Ideally, we would like to derive analytical expressions for $\overline{\tau}$ and $g(\theta)$. However, this is a non-trivial mathematical problem due to the underlying two-way diffusion process. In the appendices we outline a hybrid analytical/numerical scheme for solving the full model equations in Laplace space, assuming that the particle starts in a state bound to the absorbing wall. The hybrid scheme, which was originally developed to solve the steady-state equations \cite{Wagner17,Wagner19,Wagner22}, generates a solution in the form of an infinite sum of Hill functions,  which have to be computed numerically. The coefficients in the sum are determined from the boundary data using an infinite Neumann series.
In practice, it is necessary to truncate the infinite Neumann series to some finite order $n$, and evaluate the resulting approximation numerically. (This yields reasonable numerical accuracy in the case of the steady-state model \cite{Wagner17,Wagner19,Wagner22}.) The main outstanding mathematical issue is proving that the truncated solution converges to the exact solution in the limit $n\rightarrow \infty$.

Another difficulty of working with the full model equations is that incorporating a more general form of absorption is non-trivial compared to our formulation. This suggests that if the convergence issues could be resolved, then it would be advantageous to modify the hybrid scheme presented in the appendices to calculate $g(\theta)$ and $\overline{\tau}$ instead. This would require replacing the boundary conditions (\ref{q}) and (\ref{J}) by the totally absorbing boundary condition $p(0,\theta,t)=0$ for all $\theta \in (-\pi/2,\pi/2)$, and taking the initial condition to be of the form $p(x,\theta,0)=\delta(x)[\sigma_+\delta(\theta-\pi/2)+\sigma_-\delta(\theta+\pi/2)]$, with $\sigma_{\pm}\geq 0$ and $\sigma_-+\sigma_+=1$. The modified initial condition would now appear in the Laplace transform of equation (\ref{FP}).

\setcounter{equation}{0}
\renewcommand{\theequation}{A.\arabic{equation}}
\section*{Appendix A: Evolution equation in Laplace space}

In appendices A-C we outline the basic steps required to solve the full evolution equations in Laplace space, highlighting some of the technical difficulties. We proceed by extending the hybrid analytical/numerical method for solving the corresponding steady-state equations \cite{Lee13,Wagner17,Wagner19,Wagner22}. 
Suppose that the particle starts in the bound state $B_0$ with orientation $\theta_0$, that is, $X(0)=0$ and $\Theta(0)=\theta_0 \in {\mathcal I}_-$. Setting $\widetilde{p}(x,\theta,s)=\int_0^{\infty}\e^{-st}p(x,\theta, t)dt$, the Laplace transformed evolution equations take the form
\begin{subequations}
\begin{align}
\label{FPLT}
 D\frac{\partial^2\p(x,\theta,s)}{\partial \theta^2}&=v_0\cos\theta \frac{\partial \p(x,\theta,s)}{\partial x}+s\p(x,\theta,s),\quad x\in (0,L), \quad \theta \in [-\pi,\pi],\\
  \label{q0LT}
  D\frac{\partial^2\Q_0(\theta,s)}{\partial \theta^2}&= 
v_0\cos \theta  \p(0,\theta,s) +(s+\kappa_0)\Q_0(\theta, s)-\delta(\theta-\theta_0) ,\ \theta \in {\mathcal I}_-,\\
\label{J0LT}
v_0 \cos \theta \p(0,\theta,s)&= D\frac{\partial \Q_0(\pi/2,s)}{\partial \theta}\delta(\theta-\pi/2+\epsilon)-D\frac{\partial \Q_0(-\pi/2,s)}{\partial \theta}\delta(\theta+\pi/2-\epsilon),\\
\label{qLLT}
D\frac{\partial^2\Q_L(\theta,s)}{\partial \theta^2} &=
-v_0\cos \theta  \p(L,\theta,s) +s\Q_L(\theta,s),\ \theta \in {\mathcal I}_+,\\
\label{JLLT}
 v_0 \cos \theta \p(L,\theta,s)&=-D\frac{\partial \Q_L(\pi/2,s)}{\partial \theta}\delta(\theta-\pi/2-\epsilon)+D\frac{\partial \Q_L(-\pi/2,s)}{\partial \theta}\delta(\theta+\pi/2+\epsilon).
\end{align}
\end{subequations}
Given a solution for $\Q(\theta_0,s)$, the MFPT for absorption could be determined directly from equation (\ref{MFPT}). 

The boundary value problem (BVP) given by equations (\ref{FPLT})--(\ref{JLLT}) is a non-trivial generalization of the steady-state BVP analyzed previously in the absence of absorption \cite{Lee13,Wagner17,Wagner19}. The first step is to introduce the separable solution $\p(x,\theta,s)=X(x)\Theta(\theta,s)$ into equation (\ref{FPLT}), which yields the pair of ODEs
\begin{subequations}
\begin{align}
\label{sepx}
\frac{dX}{dx}&=\lambda X/\ell,\quad \ell =\frac{v_0}{D},\\
\frac{d^2\Theta}{d\theta^2}-\lambda \cos \theta  \Theta &=\frac{s}{D}
\Theta,\quad \Theta(\theta,s)=\Theta(\theta+2\pi,s).
\label{sepq}
\end{align}
\end{subequations}
(The separable functions $X(x)$ and $\Theta(\theta,s)$ should be distinguished from the stochastic variables $X(t)$ and $\Theta(t)$.) The first equation has the solution $\e^{\lambda X/\ell}$ for a given $\lambda$. 
The second equation for $s>0$ is related to Hill's eigenvalue equation. (The standard eigenvalue problem for Hill's equation would be $d^2 \Theta/d\theta^2+f(\theta)\Theta =\lambda \Theta$ where $f(\theta)$ is periodic \cite{Magnus12}.)  Since $\cos \theta$ is an even function of $\theta$, the eigenfunctions can be partitioned into odd and even subsets. If $\Theta(\theta)$ is an eigenfunction corresponding to an eigenvalue $\lambda$,  then $\overline{\Theta}(\theta)\equiv \Theta(\theta+\pi)$ is an eigenfunction whose eigenvalue is $-\lambda$. The proof is straightforward:
\begin{align*}
 \frac{d^2\overline{\Theta}(\theta)}{d^2\theta}&= \frac{d^2\Theta(\theta+\pi)}{d^2\theta}=\lambda\cos(\theta+\pi) \Theta(\theta+\pi) =-\lambda \cos \theta \overline{\Theta}(\theta).
\end{align*}
It can also be proven that the non-zero eigenvalues are non-degenerate \cite{Fisch80}, which motivates the following ordering of the non-zero eigenvalues:  
 $\ldots < \lambda_2  <\lambda_1 <0 < \lambda_{-1} <\lambda_{-2} <\ldots \mbox{ with } \lambda_k =-\lambda_{-k} .$
The eigenfunctions $\Theta_k(\theta,s)$ satisfy 
the orthogonality relation
 \begin{equation}
 \label{orthob}
 \int_{-\pi}^{\pi}\Theta_j(\theta,s)\Theta_k(\theta,s)\cos\theta d\theta=\delta_{i,j} \mbox{sgn}(k),\quad jk\neq 0.
 \end{equation}
In order to establish equation (\ref{orthob}), consider two distinct eigenvalues $\lambda_k$ and $\lambda_l$ with $l\neq k$. Multiply the eigenvalue equation for $\Theta_k $ by $\Theta_j$ and multiply the eigenvalue equation for $\Theta_j $ by $\Theta_k$. Subtracting the pair of equations gives
 \begin{equation}
 \Theta_j(\theta,s)  \frac{d^2\Theta_k(\theta,s)}{d\theta^2}-\Theta_k(\theta,s) \frac{d^2\Theta_j(\theta,s)}{d\theta^2}=(\lambda_k-\lambda_j) \cos \theta  \Theta_j(\theta,s) \Theta_k(\theta,s) .
 \end{equation}
 Integrating both sides with respect to $\theta$ using integration by parts and periodicity of the eigenfunctions yields equations (\ref{orthob}) for $k\neq j$. This is supplemented by
 the normalization
 $\int_{-\pi}^{\pi}\Theta_k^2(\theta)\cos \theta d\theta=\mbox{sgn}(k)$. 

The set of eigenfunctions $\{\Theta_k(\theta,s),k\neq 0\}$ with corresponding non-zero eigenvalues $\lambda_k(s)$ form a complete basis set when $s>0$. Hence, the general solution for $s>0$ has the spectral decomposition
\begin{equation}
 \p(x,\theta,s)=\sum_{k>0}c_k(s)\e^{\lambda_k(s)x/\ell}\Theta_k(\theta,s)+\sum_{k<0}c_k(s)\e^{\lambda_k(s)[x-L]/\ell}\Theta_k(\theta,s).
\label{gensolLT}
\end{equation}
There is an additional complication when $s=0$, since $\{\Theta_k(\theta,s),k\neq 0\}$ is no longer complete. The same issue arises for the steady-state solution, and is a consequence of the fact that
the eigenfunctions $\Theta_k(\theta,0)$ satisfy 
the additional orthogonality relations
\begin{equation}
\label{orthoa}
\int_{-\pi}^{\pi}\Theta_k(\theta,0)\cos\theta d\theta=0,\quad \int_{-\pi}^{\pi}\Theta_k(\theta,0)\cos^2\theta d\theta=0,\quad k\neq 0.
 \end{equation}
Equations (\ref{orthoa}) follow directly from integrating equation (\ref{sepq}) for $s=0$. However, there now exists a doubly degenerate zero eigenvalue whose eigenspace is spanned by the functions $u_0=\alpha$ and $\widehat{u}_0=\beta (x/\ell-\cos\theta)$, where $\alpha,\beta$ are constants. The latter eigenfunction is non-separable and is known as the diffusion solution. Inclusion of this additional pair of eigenfunctions generates a complete basis set, leading to a general solution of the form 
\begin{equation}
 \p(x,\theta,0)=\alpha+\beta(x/\ell-\cos \theta)+\sum_{k>0}d_k\e^{\mu_k x/\ell}\Theta_k(\theta,s)+\sum_{k<0}d_k\e^{\mu_k [x-L]/\ell}\Theta_k(\theta,s).
\label{gensolLT0}
\end{equation}

Since the FPT problem involves the small-$s$ limit, it is instructive to see how the constant and diffusion solution of the steady-state solution emerge in the limit $s\rightarrow 0$ using a perturbation argument similar to one introduced in Fisch and Kuskal (1980)\cite{Fisch80}. As we show below,
\begin{equation}
\lim_{s\rightarrow 0}\lambda_{\pm 1}(s)= 0,\quad \lim_{s\rightarrow 0}\lambda_k(s) =\mu_{k-1},\quad k>1,\quad \lim_{s\rightarrow 0}\lambda_k(s) =\mu_{k+1},\quad k<-1.
\end{equation}
We thus have the ordering 
$
 \ldots < \mu_2  <\mu_1 <0 < \mu_{-1}<\mu_{-2} <\ldots \mbox{ with } \mu_k =-\mu_{-k}.
$
It also follows that $d_{k-1}=c_k(0)$ for $k>1$ and $d_{k+1}=c_k(0)$ for $k < -1$. (It  turns out that $c_{-1}(s)$ is singular in the limit $s\rightarrow 0$.) First, non-dimensionalize by taking $D=1$. Suppose that $\lambda_{\pm 1}$ are $O(\sqrt{s})$ eigenvalues when $0<s\ll 1$. (For $D\neq 1$ we replace $s$ by $s/D$.)  Introduce the perturbation series expansion
\begin{equation}
\lambda_k(s)=\sqrt{s}\lambda_k^{(0)}+s\lambda_k^{(1)}+ \ldots,\quad \Theta_k(\theta,s)= \Theta_k^{(0)}(\theta)+\sqrt{s}  \Theta_k^{(1)}(\theta)+\ldots, \quad k=\pm 1.
\end{equation}
Substituting into equation (\ref{sepq}) and collecting terms in powers of $s$ yields a hierarchy of equations. The zeroth order and $O(\sqrt{s})$ equations are
\begin{equation}
\frac{d^2\Theta_k^{(0)}}{d\theta^2}=0,\quad \frac{d^2\Theta_k^{(1)}}{d\theta^2}=C_k\lambda_k^{(0)} \cos \theta,\quad k=\pm 1.
\end{equation} 
Hence, $\Theta_k^{(0)}=C_k$, where $C_k$ is a non-zero constant, and $\Theta_k^{(1)}=-C_k\lambda_k^{(0)}\cos \theta $. At $O(s)$ we have
\begin{align}
&\frac{d^2\Theta_k^{(2)}}{d\theta^2}=-[\lambda_k^{(0)}]^2C_k \cos^2 \theta +C_k\lambda_k^{(1)} \cos \theta +C_k.
\label{hier2}
\end{align}
Integrating both sides with respect to $\theta$ and imposing periodicity gives\begin{equation}
C_k\int_{-\pi}^{\pi} \left [1-[\lambda_k^{(0)}]^2 \cos^2 \theta \right ]d\theta =0,\quad k=\pm 1.
\end{equation}
Hence, $\lambda_{\pm 1}^{(0)}=\mp \sqrt{ 2}$.
We can now solve equation (\ref{hier2}), which shows that
\begin{equation}
 \Theta_k^{(2)}(\theta)= \frac{C_k}{4}\cos 2\theta-C_k\lambda_k^{(1)}\cos \theta+C'_k, \quad k=\pm 1,
 \end{equation}
 for some constant $C'_k$. The $O(s)$ terms $\lambda_k^{(1)}$ and $C_2$ can be determined by going to higher order. 
 
We have thus found a pair of eigenfunctions of the form
\begin{equation}
\Theta_k(\theta,s)=C_k\left \{ 1+\mbox{sgn}(k)\sqrt{ 2s}\cos \theta +\frac{s}{4}\frac{\cos 2\theta}{4}-s\lambda_k^{(1)}\cos \theta\right \}+o(s), \quad k=\pm 1,
\end{equation}
where we have combined the constant term $sC_k'$ with $C_k$. Substituting into the $k=\pm 1$ terms of the general solution (\ref{gensolLT}) and absorbing the factor $C_k$ into the corresponding coefficient $c_k(s)$, we find that
\begin{align}
 &c_k(s)\e^{\lambda_k(s)x/\ell}\Theta_k(\theta,s)\nonumber \\
 &=c_k(s)\exp\left ([\sqrt{s}\lambda_k^{(0)}+s\lambda_k^{(1)}+ \ldots]x/\ell\right )\left [\Theta_k^{(0)}(\theta)+\sqrt{s}  \Theta_k^{(1)}(\theta)+\ldots \right ]\nonumber\\
&=c_k(s)\left [1+\sqrt{2s}\left (\frac{x}{\ell}+\mbox{sgn}(k)\cos \theta\right )+O(s)\right ].
 \label{ckk}
\end{align}
Taking $c_1(s) =\mbox{constant}+O(\sqrt{s})$ recovers the constant solution in the limit $s\rightarrow 0$.This can then be used to subtract out the term $c_{-1}(s)$ on the right-hand side of equation (\ref{ckk}) for $k=-1$. Taking $c_{-1}(s)\sim 1/\sqrt{s}$, then generates non-separable diffusion solution $x/\ell-\cos \theta$ in the limit $s\rightarrow 0$.

\setcounter{equation}{0}
\renewcommand{\theequation}{B.\arabic{equation}}
\section*{Appendix B: Matching the boundary conditions at the walls}

Let us first consider the simplified boundary conditions
\begin{equation}
 \p(0,\theta,s)=v_+(\theta,s) \mbox{ for } \cos \theta >0,\quad \p(L, \theta,s)= v_-(\theta,s) \mbox{ for } \cos \theta <0.
 \label{bw}
\end{equation}
Substituting the general solution (\ref{gensolLT}) into equations (\ref{bw}) yields the pair of equations
\begin{subequations}
\begin{align}
 &v_+(\theta,s)=f_+(\theta,s)\equiv \sum_{k>0}c_k(s)\Theta_k(\theta,s)+\sum_{k<0}c_k(s)\e^{-\lambda_k(s)L/\ell}\Theta_k(\theta,s) \mbox{ for } \cos \theta >0, \label{vfp}
\\
 &v_-(\theta,s) =f_-(\theta,s)\equiv \sum_{k>0}c_k(s)\e^{\lambda_k(s)L/\ell} \Theta_k(\theta,s)+\sum_{k<0}c_k(s)\Theta_k(\theta,s) \mbox{ for } \cos \theta <0. 
\label{vfm}
\end{align}
\end{subequations}
It is useful to rewrite equations ({\ref{vfp}) and (\ref{vfm}) for fixed $s$ by introducing the space ${\mathcal H}$ of continuous piecewise, twice-differentiable functions of $\theta$ and defining $v(\cdot,s),f(\cdot,s)\in {\mathcal H}$ with 
\begin{align}
  v(\theta,s)=\left \{\begin{array}{cc} v_+(\theta,s), &  \cos \theta >0\\
 v_-(\theta,s),&\ \cos \theta <0\end{array}\right . ,\quad
f(\theta,s)=\left \{\begin{array}{cc} f_+(\theta,s), &  \cos \theta >0\\
 f_-(\theta,s),&\ \cos \theta <0\end{array}\right . .
 \end{align}
 The boundary condition can then be written in the more compact form $f(\theta,s)=v(\theta,s)$ for all $\theta\in (-\pi,\pi)$. It can be shown that the eigenfunctions $\{\Theta_k(\theta,s),k\neq 0\}$ form a complete basis set for ${\mathcal H}$ \cite{Fisch80,Beals83,Beals85,Wagner19}. This means that we can set $v(\theta,s)=\sum_{k\neq 0}v_k^{(0)}(s)\Theta_k(\theta,s)$ with
 \begin{equation}
 \label{vk}
v_k^{(0)}(s)\equiv \mbox{sgn}(k)\int_{-\pi}^{\pi} v(\theta,s) \Theta_k(\theta,s)\cos \theta d\theta.
\end{equation}
Equation (\ref{vk}) follows from the orthogonality relation (\ref{orthob}).  

 Following Ref. \cite{Wagner19}, we now introduce two sets of projection operators. The first pair $\calQ_{\pm}$ projects an element $u(\cdot,s)\in {\mathcal H}$ onto functions restricted to the domains $\theta \in {\mathcal I}_{\pm}$:
 \begin{equation}
  \calQ_+ u(\theta,s)=\left \{\begin{array}{cc} u_+(\theta,s), &  \cos \theta >0\\
 0 &\ \cos \theta <0\end{array}\right .,\quad \calQ_- u(\theta,s)=\left \{\begin{array}{cc}0, &  \cos \theta >0\\
  u_-(\theta,s) &\ \cos \theta <0\end{array}\right . .
  \end{equation}
In order to define the second pair of projection operators $\calP_{\pm}$, we expand an arbitrary element $u(\cdot,s)\in {\mathcal H}$ as $u(\theta,s) =\sum_{k\neq 0}u_k(s) \Theta_k(\theta,s)$ and set
\begin{align}
 \calP_+ u(\theta,s)= \sum_{k >0}u_k(s) \Theta_k(\theta,s),\quad \calP_- u(\theta,s)= \sum_{k <0}u_k(s) \Theta_k(\theta,s).
\end{align}
 We now note that functions $f_{\pm}(\theta,s)$ in equations (\ref{vfp}) and (\ref{vfm}) can be rewritten as
 \begin{equation}
 f_+(\theta,s)=[\calP_++\calP_-{\mathcal M}_L]F(\theta,s),\quad f_-(\theta,s)=[\calP_+{\mathcal M}_L+\calP_-]F(\theta,s),
\end{equation}
where 
\begin{equation}
 F(\theta,s):=\sum_{k\neq 0} c_k(s)\Theta_k(\theta,s),\quad {\mathcal M}_LF(\theta,s):=\sum_{k\neq 0} c_k(s)\e^{-|\lambda_k(s)|L/\gamma}\Theta_k(\theta,s).
 \end{equation}
It follows that the boundary condition becomes
\begin{align}
v(\theta,s)&=\calQ_+f_+(\theta,s)+\calQ_-f_-(\theta,s)\nonumber \\
&=\left \{\calQ_+[\calP_++\calP_-{\mathcal M}_L]+\calQ_-[\calP_+{\mathcal M}_L+\calP_-]\right \}F(\theta,s)\nonumber\\
&={\mathcal V}F(\theta,s) +{\mathcal W}{\mathcal M}F(\theta,s),
\end{align}
where
${\mathcal V}=\calQ_+\calP_++\calQ_-\calP_-,\quad {\mathcal W}=\calQ_+\calP_-+\calQ_-\calP_+$.
Using the operator identity ${\mathcal V}+{\mathcal W}=(\calQ_++\calQ_-)(\calP_++\calP_-)=I$, we obtain the result
\begin{equation}
v(\theta,s)=(I-{\mathcal W}_L)F(\theta,s),\quad {\mathcal W}_L={\mathcal W}-{\mathcal W}{\mathcal M}_L,
\end{equation}
which can be inverted in terms of a Neumann series \cite{Wagner19}
\begin{equation}
\label{Fsum}
F(\theta,s)=\sum_{n=0}^{\infty} {\mathcal W}_L^n v(\theta,s).
\end{equation}

A non-trivial issue is whether or not the infinite series representation of $F(\theta,s)$ converges. In terms of the $L^2$ inner product, this is equivalent to the condition  $\|{\mathcal W}_L\|<1$. As discussed by Wagner {\em et al.} \cite{Wagner19}, the norm of ${\mathcal W}_L$ is difficult to estimate. However, in practice, one can establish convergence numerically by restricting the Hilbert space ${\mathcal H}$ to the space ${\mathcal H}_N=\mbox{span}\{\Theta_k,|k\leq N\}$, that is, the space spanned by the first $2N$ eigenfunctions ordered by the magnitude of their corresponding eigenvalues. Modifying the definitions of the projection operators accordingly, one finds that $\|{\mathcal W}_N\|<1$ for values of $N$ up to $O(10^3)$ with an asymptote suggesting that $\lim_{N\rightarrow \infty} \|{\mathcal W}_N\| <1$ (see Fig. 1 of Wagner {\em et al.} \cite{Wagner19}). It remains to justify approximating solutions by restricting to the space ${\mathcal H}_N$. This is valid provided that the BVP defined by equations (\ref{sepx}), (\ref{sepq}) and (\ref{bw}) has solutions that are sufficiently smooth and slowly varying. Since eigenfunctions with larger eigenvalues are faster varying, it follows that they do not contribute significantly. Assuming that the Neumann series (\ref{Fsum}) is convergent, one can generate a sequence of approximate analytic solutions along analogous lines to Wagner {\em et al.} \cite{Wagner19}. Let $F^{(n)}(\theta,s)$ denote the approximation obtained by truncating the series solution at the $n$th term. It immediately follows that the zeroth order solution is 
$
F^{(0)}(\theta,s)=\sum_{k\neq0} v_k^{(0)}(s)\Theta_k(\theta,s).
$
At the next level of approximation, the contribution on the right-hand side of equation (\ref{Fsum}) is
\begin{align}
 {\mathcal W}_L v(\theta,s) =[\calQ_+\calP_-+\calQ_-\calP_+]v(\theta,s)=\left \{\begin{array}{cc} \sum_{k<0} v_k^{(0)}(s)\Theta_k(\theta,s), &  \cos \theta >0\\
\sum_{k>0} v_k^{(0)}(s)\Theta_k(\theta,s),& \cos \theta <0\end{array}\right . .
 \end{align}
Note that that the eigenfunctions $\{\Theta_k(\theta,s),k < 0\}$ span the set of functions restricted to the domain $\cos \theta >0$ and the subset $\{\Theta_k(\theta,s),k > 0\}$ span the set of functions restricted to the domain $\cos \theta <0$; this is known as the half-range completeness property \cite{Fisch80,Beals83,Beals85}. Rewriting ${\mathcal W}_L v(\theta),s $ as 
\begin{equation}
{\mathcal W}_L v(\theta,s)= \sum_{k\neq 0} v^{(1)}_{k}(s) \Theta_k(\theta,s),\quad  v_k^{(1)}(r)=\int_{-\pi}^{\pi} {\mathcal W}V_r(\theta) \Theta_k(\theta,r)\cos \theta d\theta
\end{equation}
with $\theta\in [0,2\pi]$ leads to the next level approximation
$
 F^{(1)}(\theta,s)= \sum_{k\neq 0} [v_k^{(0)}(s)+v_{k}^{(1)}(s]\Theta_k(\theta,s) $. 
Iterating the procedure generates an approximation to arbitrary order $n$ with $c_k(s)\approx \sum_{j=0}^n v_k^{(j)}(s)$.

So far the analysis has neglected the tumbling dynamics and absorption at the walls, as determined by the bound state probability densities $\Q_0(\theta,s)$ and $\Q_L(\theta,s)$. If the latter were known explicitly then one could determine the functions $v_{\pm}(\theta,s)$ from the boundary conditions (\ref{J0LT}) and (\ref{JLLT}):
\begin{subequations}
\begin{align}
\label{AB1}
 &  v_+(\theta,s)= \frac{A_0(s)}{\ell \cos \theta} \delta(\theta-\pi/2+\epsilon)+\frac{B_0(s)}{\ell \cos \theta}\delta(\theta+\pi/2-\epsilon) ,\\
& v_-(\theta,s) = -\frac{A_L(s)}{\ell \cos \theta} \delta(\theta-\pi/2-\epsilon)-\frac{B_L(s)}{\ell \cos \theta}\delta(\theta+\pi/2+\epsilon) , 
\label{AB2}
\end{align}
\end{subequations}
where
\begin{align}
\label{defA}
A_{0,L}(s)&:=\frac{\partial \Q_{0,L}(\pi/2,s)}{\partial \theta},\
B_{0,L}(s):=-\frac{\partial \Q_{0,L}(-\pi/2,s)}{\partial \theta}.
\end{align}
Unfortunately, the probability densities $\Q_0(\theta,s)$ and $\Q_L(\theta,s)$ are only defined implicitly, since the remaining pair of boundary conditions (\ref{q0LT}) and (\ref{qLLT}) depend on $\p(0,\theta,s)$ and $\p(L,\theta,s)$.
Hence, the unknown coefficients $A_{0,L}(s)$ and $B_{0,L}(s)$ have to be determined self-consistently.
(This differs significantly from the steady-state analysis, which is relatively straightforward to deal with since there is a balance of fluxes at both ends \cite{Wagner17}.)

\setcounter{equation}{0}
\renewcommand{\theequation}{C.\arabic{equation}}
\section*{Appendix C: Calculation of the coefficients $A_{0,L}(s)$ and $B_{0,L}(s)$}
First, consider equation (\ref{q0LT}). This can be solved in terms of the Dirichlet Green's function $G_0$ of the modified Helmholtz equation on ${\mathcal I}_-$. Note that $G_0(\theta,s|\theta_0)=\trho(\theta+\pi,s|\theta_0+\pi)$ with $\trho$ defined in equation (\ref{aFPLT}). The solution takes the form
\begin{equation}
  \Q_0(\theta,s)=-v_0 \int_{{\mathcal I}_-}G_0(\theta,s+\kappa_0|\theta')\cos \theta'  \p(0,\theta',s)d\theta' -G_0(\theta,s+\kappa_0|\theta_0).
\label{q0LT2}
\end{equation}
Substituting the general solution (\ref{gensolLT}) into equation (\ref{q0LT2}) gives
\begin{align}
  \Q_0(\theta,s)&=-v_0\left [ \sum_{k>0}c_k(s)\Phi_{0,k}(\theta,s)+\sum_{k<0}c_k(s) \e^{-\lambda_k(s)  L/\ell}\Phi_{0,k}(\theta,s) \right ] \nonumber \\
  &\quad -G_0(\theta,s+\kappa_0|\theta_0),
\label{0solQ0}
\end{align}
with
\begin{equation}
\Phi_{0,k}(\theta,s):= \int_{{\mathcal I}_-}G_0(\theta,s+\kappa_0|\theta')\cos \theta'  \Theta_k(\theta',s)d\theta'  .
 \end{equation}
 Differentiating both sides with respect to $\theta$ and using the definitions of $A_0$ and $B_0$ in (\ref{defA}) yields
 \begin{align}
 \label{A0}
  A_0(s)&=-v_0\left [\sum_{k>0}c_k(s)\Psi^+_{0,k}(s)+\sum_{k<0}c_k(s) \e^{-\lambda_k(s) L/\ell}\Psi^+_{0,k}(s)\right ] -\partial_{\theta}G_0(\pi/2,\kappa_0|\theta_0),\\
  B_0(s)&=v_0\left [\sum_{k>0}c_k(s) \Psi^-_{0,k}(s)+\sum_{k<0}c_k(s) \e^{-\lambda(s)_k L/\ell}\Psi^-_{0,k}(s) \right ] \nonumber \\
  &\quad +\partial_{\theta}G_0(-\pi/2,s+\kappa_0|\theta_0).
  \label{B0}
 \end{align}
 with $\Psi^{\pm}_{0,k}(s):=\partial_{\theta}\Phi_{0,k}(\pm \pi/2)$.
A similar analysis can be applied to equation (\ref{qLLT}) using the corresponding Dirichlet Green's function $G_L$ on ${\mathcal I}_+$, where $G_L(\theta,s|\theta_0)=\trho(\theta,s|\theta_0)$:
\begin{equation}
 \Q_L(\theta,s)=v_0 \int_{{\mathcal I}_+}G_L(\theta,s|\theta')\cos \theta'  \p(L,\theta',s)d\theta' .
\label{qLLT2}
\end{equation}
Substituting the general solution (\ref{gensolLT}) into equation (\ref{qLLT2}), we have
\begin{align}
  \Q_L(\theta,s)&=v_0\left [\sum_{k>0}c_k(s)\e^{\lambda_k(s) L/\ell} \Phi_{L,k}(\theta,s)+\sum_{k<0}c_k(s) \Phi_{L,k}(\theta,s) \right ],
\label{0solQL}
\end{align}
with
\begin{equation}
\Phi_{L,k}(\theta,s) :=\int_{{\mathcal I}_+}G_L(\theta,s|\theta')\cos \theta'  \Theta_k(\theta',s)d\theta'  .
 \end{equation}
 Differentiating both sides with respect to $\theta$ and using the definitions of $A_L$ and $B_L$ in (\ref{defA} yields
 \begin{align}
 \label{AL}
  A_L(s)&=v_0\left [\sum_{k>0}c_k(s)\e^{\lambda_k(s)  L/\ell}\Psi^+_{L,k}(s)+\sum_{k<0}c_k(s) \Psi^+_{L,k}(s)\right ],\\
   B_L(s)&=-v_0\left [\sum_{k>0}c_k(s)\e^{\lambda_k(s)  L/\ell}\Psi^+_{L,k}(s)+\sum_{k<0}c_k(s) \Psi^+_{L,k}(s) \right ],
   \label{BL}
 \end{align}
 with $\Psi^{\pm}_{L,k}(s):=\partial_{\theta}\Phi_{L,k}(\pm \pi/2)$. Finally, the coefficients $c_k(r)$ can be expressed as linear combinations of the coefficients $A_{0,L}(s)$ and $B_{0,L}(s)$ using the sequence of approximations based on the Neumann series solution (\ref{Fsum}). We find that
 at order $n$
\begin{align}
c_k(r)
 \approx \frac{\mbox{sgn}(k)}{\ell}
\bigg \{[A_0(s)-A_L(s)]\sum_{j=0}^n\Lambda^{(j)}_{k,+}(s)+[B_0(s)-B_L(s)] \sum_{j=0}^n{\Lambda}^{(j)}_{k,-}(s)\bigg \}.
 \label{vk2n}
\end{align}
In particular,  the lowest order contributions are $\Lambda^{(0)}_{k,\pm}(r)=\Theta_k(\pm \pi/2,r)$, and 
\begin{align}
\Lambda_{k,\pm}^{(1)}(s)&=\sum_{l<0}M^{+}_{kl}(s)\Theta_l(\pm \pi/2,s)+ \sum_{l>0}M^{-}_{kl}(s)\Theta_l(\pm \pi/2,s),\end{align}
with
\begin{equation}
M^{\pm}_{kl}(s)=\int_{{\mathcal I}_{\pm}}\Theta_l(\theta,s)\Theta_k(\theta,s)\cos \theta d\theta.
\end{equation}
Finally, substituting equation (\ref{vk2n}) into equations (\ref{A0}), (\ref{B0}), (\ref{AL}) and (\ref{BL}) yields a linear inhomogeneous system of equations that determines the coefficients $A_{0,L}(s)$ and $B_{0,L}(s)$ at the given level of approximation.

\end{document}